\documentclass[twocolumn,showpacs,preprintnumbers,amsmath,amssymb]{revtex4}
\usepackage{graphicx}
\usepackage{dcolumn}
\usepackage{bm}

\usepackage{mathptmx}

\begin{document}

\title{Influence of the Ground-State Topology on the
Domain-Wall Energy in the Edwards-Anderson $\pm J$ Spin Glass
Model}

\author{F. Rom\'a}
\affiliation{Centro At{\'{o}}mico Bariloche and CONICET, R8402AGP
San Carlos de Bariloche, R\'{\i}o Negro, Argentina}
\author{S. Risau-Gusman}
\affiliation{Centro At{\'{o}}mico Bariloche and CONICET, R8402AGP
San Carlos de Bariloche, R\'{\i}o Negro, Argentina}
\author{A.J. Ramirez-Pastor}
\author{F. Nieto}
\affiliation{Departamento de F\'{\i}sica, Universidad Nacional de
San Luis and CONICET, Chacabuco 917, D5700BWS San Luis, Argentina}
\author{E.E. Vogel}
\affiliation{Departamento de F\'{\i}sica, Universidad de La
Frontera, Casilla 54-D, Temuco, Chile}

\begin{abstract}
We study the phase stability of the Edwards-Anderson spin-glass
model by analyzing the domain-wall energy. For the bimodal $\pm J$
distribution of bonds, a topological analysis of the ground state
allows us to separate the system into two regions: the {\em
backbone} and its {\em environment}. We find that the
distributions of domain-wall energies are very different in these
two regions for the three dimensional (3D) case. Although the
backbone turns out to have a very high phase stability, the
combined effect of these excitations and correlations produces the
low global stability displayed by the system as a whole. On the
other hand, in two dimensions (2D) we find that the surface of the
excitations avoids the backbone. Our results confirm that a narrow
connection exists between the phase stability of the system and
the internal structure of the ground-state. In addition, for both
3D and 2D we are able to obtain the fractal dimension of the
domain wall by direct means.
\end{abstract}


\date{\today}

\maketitle

The spin glass state has been studied extensively during the last
thirty years, but the role played by low energy excitations is
still a matter of debate. These excitations are crucial to
understand the nature of the ordering of the spin glass phase.
Most studies have focused on the predictions of two theories: the
replica-symmetry breaking (RSB) picture \cite{Parisi} and the
droplet picture \cite{Fisher1}. RSB, rigorously true
for the Sherrington-Kirkpatrick model of spin glasses, predicts
that there are excitations which involve flipping a finite
fraction of the spins and, in the thermodynamic limit, cost only a
finite amount of energy. The fractal dimension of the surface of
these excitations, $d_s$, is expected to be equal to the space
dimension, $d$. On the other hand, in the droplet picture the
lowest energy excitations of length $L$ have $d_s<d$ and typically
cost an energy of order $L^\theta$ ($\theta$ is known as stiffness
exponent). Thus, contrary to RSB, the droplet picture predicts that
excitations involving a finite fraction of spins cost an infinite
amount of energy in the thermodynamic limit.

The exponent $\theta$ plays a central role in this debate. It is
usually calculated by using the concept of defect energy, $\Delta
E=E^a-E^p$, which is the difference between the ground-state (GS)
energies for antiperiodic ($E^a$) and periodic ($E^p$) boundary
conditions, in one of the directions of a $d$-dimensional system
of linear size $L$. In ferromagnetic systems, $\Delta E  \sim
L^\theta$, with $\theta = d_s = d-1$, because the induced defect
is a ($d-1$)-dimensional domain wall with {\em all} its bonds
frustrated. For spin glasses, the average over the distribution of
bonds (denoted by $\left[ ... \right]$) must be taken and the
scaling ansatz becomes
\begin{equation}
\left[ \vert \Delta E \vert \right] \sim L^\theta. \label{def}
\end{equation}
Assuming that, because of frustration, the defect energy
is the sum of many correlated terms of different signs, Fisher and
Huse \cite{Fisher1} have shown that for spin glasses $\theta \leq
\frac{d-1}{2}$.

It is well known that the Edwards-Anderson (EA) $\pm J$ model
\cite{EA} has a degenerate GS. In addition, it has been shown
\cite{Toulouse01,Toulouse02,Barahona} that, for each realization
of the disorder in two dimensions (2D), there are bonds which are
either {\em always} satisfied or {\em always} frustrated in all
the GSs. These bonds define the {\em Rigid Lattice} (RL), or {\em
backbone} of the system. Spins connected by it are called {\em
solidary} spins (the rest are denoted {\em non-solidary}).
Numerical studies in three dimensions (3D) of a similar structure,
the {\em Diluted Lattice} \cite{Ramirez2} (defined by the
satisfied bonds of the RL), seem to confirm the existence of the
RL in the thermodynamic limit.

Recently, an interesting connection has been found between the RL
and both thermodynamic and dynamic properties
\cite{Roma01,Ricci,Roma02}. For example, in Ref.~\cite{Roma02} the
{\em slow} and {\em fast} degrees of freedom associated to the
out-of-equilibrium dynamics of the 2D EA $\pm J$ model were shown
to be closely related to the solidary and non-solidary spins,
respectively. In this context, one expects that the contributions
of the backbone and its environment on the domain-wall energy will
be rather different. The present letter represents a step in that
direction.

In this work we show that in 3D a significant portion of the
domain wall is contained in the backbone. Moreover, in this region
we find that the defect energy is positive and has a strong size
dependence, similar to what happens in a ferromagnetic system. The
rest of the system has a negative defect energy and could
therefore be considered as an excited phase. The sum of these two
defect energies results in the cancellation effect responsible for
the weak size dependence observed for the domain-wall energy
\cite{Hartmann3,Cartery}. On the other hand, we find that in 2D
the portion of the domain wall inside the backbone is vanishing.
We have also obtained the fractal dimensions of domain wall for
both 2D and 3D.


We start by considering the Hamiltonian of the EA model for spin
glasses \cite{EA} on square and cubic lattices,
\begin{equation}
H = \sum_{( i,j )} J_{ij} \sigma_{i} \sigma_{j}, \label{ham}
\end{equation}
where $\sigma_i = \pm 1$ is the spin variable and $( i,j )$
indicates a sum over nearest neighbors. The coupling constants are
independent random variables chosen from a $\pm J$ bimodal
distribution.


\begin{table}
\caption{\label{t1} Parameters of the simulation for each lattice
size $L$ in 3D. $N_T$ is the number of temperatures used in the
parallel tempering, chosen inside the interval $T_{min}$ to
$T_{max}$ (temperatures in units of $J/k_B$), MCS is the number of
Monte Carlo steps needed to reach the GS and $N_S$ is the number
of samples used. $N_{GS}$ is the mean number of GSs used to
approximate the average ($\langle ... \rangle_{GS}$) for the
antiperiodic system.}
\begin{ruledtabular}
\begin{tabular}{ccccccc}
$L$&$N_T$&$T_{min}$&$T_{max}$&MCS & $N_S$ & $N_{GS}$ \\
\hline
$3$ & $20$ & $0.1$ & $1.6$ & $2 \cdot 10^2$ & $1 \cdot 10^4$ & $3 \cdot 10^4$ \\
$4$ & $20$ & $0.1$ & $1.6$ & $2 \cdot 10^2$ & $1 \cdot 10^4$ & $5 \cdot 10^4$ \\
$5$ & $30$ & $0.1$ & $1.6$ & $2 \cdot 10^3$ & $6 \cdot 10^3$ & $2 \cdot 10^5$ \\
$6$ & $30$ & $0.1$ & $1.6$ & $2 \cdot 10^5$ & $3 \cdot 10^3$ & $4 \cdot 10^5$ \\
$7$ & $40$ & $0.1$ & $1.6$ & $2 \cdot 10^5$ & $1 \cdot 10^3$ & $1 \cdot 10^6$ \\
$8$ & $40$ & $0.1$ & $1.6$ & $2 \cdot 10^6$ & $3 \cdot 10^2$ & $2 \cdot 10^6$ \\
\end{tabular}
\end{ruledtabular}
\end{table}

For 3D lattices and periodic boundary conditions, we determine the
RL by using an improvement of the algorithm introduced in Ref.
\cite{Ramirez1}, where parallel tempering \cite{Hukushima} has
been implemented for reaching the GS. The present scheme, called
Rigid Lattice Searching Algorithm (RLSA), allows to obtain true
GSs up to $L=12$, where $L$ is the lattice size. However, the
maximum size used in this work is $L=8$, because to obtain the RL,
the GS must be reached $3N$ ($N=L^d$) times. An important point is
that the set of parameters used in the parallel tempering (see
Table I) has not been chosen to equilibrate the system, but to
reach quickly a GS configuration. To check this, for each lattice
size we have compared the average GS energy per spin, $e_0$,
calculated from our algorithm, with the value reported in the
literature \cite{Hartmann1,Hartmann2}. For example, for $L=8$ we
obtain $e_0=-1.780(1)$, in good agreement with the value
$e_0=-1.7802(5)$ reported in Ref. \cite{Hartmann2}.

For 2D lattices, we have used a different algorithm. It is well
known \cite{Hartmann1} that the problem of finding the GS for a 2D
lattice with at least one free boundary condition can be mapped to
a minimum weighted perfect matching problem, for which very
efficient algorithms exist. To implement the RLSA we have used one
of these routines, which has allowed us to calculate the RL up to
$L=100$.

As mentioned above, the RL is formed by bonds ({\em rigid bonds})
which are either always satisfied or always frustrated in all the
GSs. The remaining bonds, called {\em flexible bonds}, form the
{\em flexible lattice} (FL). This allows us to write the
Hamiltonian (\ref{ham}) as: $H= H_r+ H_f$. Subindex $r$ ($f$)
refers to the Hamiltonian restricted to only rigid (flexible)
bonds.

To calculate the defect energy we write the GS energy of a
particular sample ($p$), as
\begin{equation}
E^p= E^p_r+ E^p_f. \label{Ep}
\end{equation}
Note that, although the GS is degenerated, $E^p$, $E^p_r$ and
$E^p_f$ remain constants on all configurations of the GS. Next, a
new sample is generated ($a$), by introducing antiperiodic
boundary conditions in one direction. Its energy can be written as
\begin{equation}
E^a= \langle E^a_r \rangle_{GS} + \langle E^a_f \rangle_{GS},
\label{Ea}
\end{equation}
where the subindexes $r$ and $f$ correspond to the restriction of
the Hamiltonian to the bonds that form the RL and FL of the {\em
periodic} system, and $\langle ... \rangle_{GS}$ denotes an
average over all the GSs of the {\em antiperiodic} system
\cite{NGS}. Notice that this average is necessary, because now
$E^a_r$ and $E^a_f$ are {\em not} constants in all the GSs. Using
Eqs. (\ref{Ep}) and (\ref{Ea}), we write the defect energy as
\begin{equation}
\Delta E= \Delta E_r+ \Delta E_f, \label{def}
\end{equation}
where $ \Delta E_r = \langle E^a_r \rangle_{GS} -E_r^p $ and $
\Delta E_f = \langle E^a_f \rangle_{GS} - E_f^p $.

\begin{figure}
\includegraphics[width=7.6cm,clip=true]{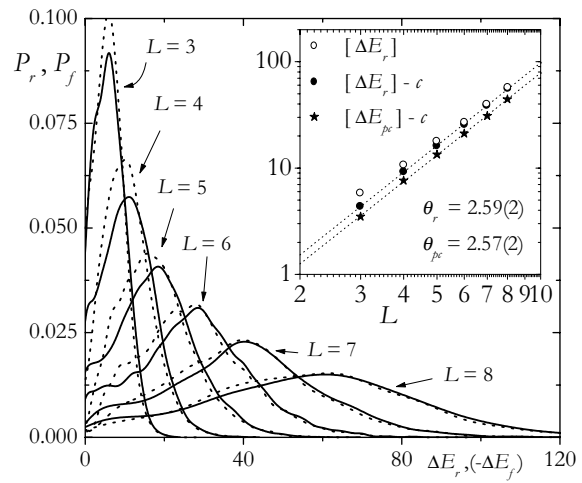}
\caption{\label{figure1} Distribution $P_r$ (full line) and $P_f$
(dotted line) for 3D. The energy is given in units of $J$.  The
inset shows the scaling of $\left[\Delta E_r \right]$,
$\left[\Delta E_r \right]-c$ \  and $\left[\Delta E_m \right]-c$.
Error bars are smaller than the symbols.}
\end{figure}

In 3D, we have measured $\left[ \vert \Delta E \vert \right]$ for
each lattice size (due to symmetry arguments, $ \left[ \Delta E
\right]=0$) and we have obtained a finite-size dependence with a
small stiffness exponent $\theta \approx 0.2$, in agreement with
the result of Refs. \cite{Hartmann3,Cartery}. This is seen as an
evidence of the existence of a finite critical temperature. We
have also measured the distribution of $\vert \Delta E \vert$. For
the sizes considered, this function extends up to $\vert \Delta E
\vert=12J$ (the values of the defect energy are multiple of $4J$).
In contrast, for the 3D ferromagnetic Ising model, the defect
energy takes the value $128J$ for $L=8$. The small values obtained
for the EA model indicate the presence of large sets of frustrated
or satisfied bonds that compensate each other.

The main point of this article is that these sets can be very
clearly related to the topology of the system. This can be seen on
Fig.~\ref{figure1}, which shows the distributions of the
contributions of the RL, $P_r (\Delta E_r )$, and the FL, $P_f
(\Delta E_f )$, to the domain-wall energy. Two features stand out.
On the one hand, $\Delta E_r$ is almost always positive. Only a
small fraction of samples have $\Delta E_r <0$, and this fraction
decreases with $L$. Our measurements indicate that on the portion
of the domain wall that crosses the RL, the fraction of
excitations (rigid bonds always satisfied that appear frustrated
in the GS of the antiperiodic system) seems to tend to 0.72. On
the other hand, $P_r$ is a broad distribution and extends up to
high values of $\Delta E_r$. In addition, our simulations show
that the distribution of $\Delta E_r$ seems to be very broad also
for {\em fixed} values of $| \Delta E |$. For example, for $L=8$,
we have found samples with $\Delta E_r =-\Delta E_f =123.59J$ or
$\Delta E_r =-\Delta E_f =11.63J$, both corresponding to $\Delta
E=0$.

The inset in Fig.~\ref{figure1} shows the scaling of $\left[\Delta
E_r \right]$ with $L$. It tends towards a power law behavior of
the form
\begin{equation}
\left[ \Delta E_r \right] \sim L^{\theta_r}. \label{defRL}
\end{equation}
To determine $\theta_r$, we have fitted the data with the function
$\left[ \Delta E_r \right] = c+b L^{\theta_r}$ (which is the
simplest correction to scaling). We obtain a good fit for $c=3/2$
and $\theta_r=2.59(2)$. The inset in Fig.~\ref{figure1} shows the
scaling of $\left[\Delta E_r \right]-c$ with $L$. Notice that the
exponent $\theta_r$ is an order of magnitude bigger than the usual
stiffness exponent $\theta$ ($\approx 0.2$ for 3D).

The behavior of $P_f (\Delta E_f)$ is very similar to the one
observed in $P_r$, but now $\Delta E_f$ is always negative (see
Fig.~\ref{figure1}). The vanishing of the average domain-wall
energy implies that $-\left[\Delta E_f \right]$ follows a power
law with the same exponent as $\left[\Delta E_r \right]$.

The previous results indicate that the proposal of separating the
system in two regions is not trivial: the sections of the
domain-wall energy with positive (negative) sign, prevail on RL
(FL). However, the system has a small but positive stiffness
$\theta$, because on average the defect energy $\left[\Delta E_r
\right]$ overcomes the defect energy $\left[\Delta E_f \right]$.

Our numerical results also allow us to infer that the exponent
$\theta_r$ is equal to the fractal dimension of domain wall,
$d_s$. To justify this conjecture, let us consider the topological
characteristics of the RL. Our simulations up to $L=8$ indicate
that the RL consists mainly of a compact {\em percolation cluster}
(PC) \cite{TrabPrep}. The fraction of the RL that corresponds to
this cluster converges to $0.78$. The fact that the distributions
$P_r$ are vanishing for negative $\Delta E_r$ shows that
compensation effects are not important in the RL. Assuming that
the same happens in the PC and that it behaves as a natural box
containing (on average) always the same fraction of domain wall as
the whole RL, it is natural to conjecture that the area of the
domain wall inside the RL follows a power law $L^{d_s}$.
Therefore, $d_s \approx \theta_r=2.59(2)$, which agrees with the
values reported in the literature for the EA model with Gaussian
distributed couplings \cite{Palassini1}.

To refine this, measurements were carried out separating the
system into two new regions: the PC and its environment (now, FL
plus small clusters of RL). Then, the defect energy can be written
as $\Delta E= \Delta E_{pc}+ \Delta E_e$, where subindexes $pc$
and $e$ refer to the defect energy of PC and its environment,
respectively. If we assume again a power law, $\left[ \Delta
E_{pc} \right] \sim L^{\theta_{pc}}$, we obtain
$\theta_{pc}=2.57(2)$, which gives $d_s=\theta_{pc}=2.57(2)$,
consistent with the value quoted above. The inset in
Fig.~\ref{figure1} shows the scaling of $(\left[\Delta E_{pc}
\right]-c)$ with $L$ for $c=3/2$.

The picture in 2D is very different. The main problem is that,
even though the RL spans a significant portion of the square
lattice, it does not percolate \cite{Barahona,TrabPrep}. It
consists instead of a large number of relatively small islands.
The analysis of the difference between periodic and antiperiodic
systems reveals that the number of bonds in the RL that belong to
the domain wall grows as $\sim L^{0.6}$. But the size of the
domain wall is necessarily larger than $L$, which implies that the
fraction of it that crosses the RL is vanishing with $L$.
Geometrically, what is happening is that the domain wall crosses
the sample avoiding the RL islands (notice that this is further
evidence that the RL does not percolate). Consequently, in the
following we shall use a different strategy for studying the
influence of the backbone on the domain-wall energy in the 2D
case.

Thus, for each sample, the domain wall depends on the pair of GSs
(one GS for the periodic and another for the antiperiodic system)
that are being compared. However, with the hope of capturing its
relevant properties, for each sample we have picked a single
random pair of GSs. But even for a pair of fixed GSs the
determination of the domain wall is not trivial. It can be shown
\cite{TrabPrep} that there are some sets of bonds of the periodic
system that appear flipped (in the sense that their satisfied
bonds become frustrated and viceversa) in the antiperiodic system
but do not contribute to the energy change. Notice that this
implies that they can also appear flipped in other GSs of the
periodic system. Therefore, one must weed out these sets of bonds
to get the correct domain wall. The plaquette picture
\cite{Toulouse} provides the best framework for this.

\begin{figure}
\includegraphics[width=7.6cm,clip=true]{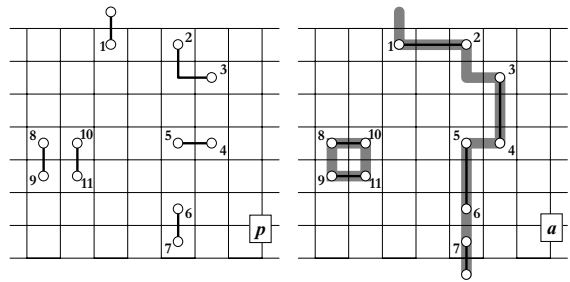}
\caption{\label{figure2} Matchings corresponding to the ground
states of a 2D sample with periodic (left) and antiperiodic
(right) boundary conditions on the horizontal direction. Points
represent frustrated plaquettes. The grey line crossing the sample
is the resulting domain wall.}
\end{figure}

The 2D lattice can be seen as a set of squares, called plaquettes,
bounded by four bonds. If an odd number of these bonds are
frustrated, the plaquette is called frustrated. It has been shown
that plaquettes are frustrated if and only if there is an odd
number of ferromagnetic bonds in the boundary
\cite{Toulouse,Bieche}. Every spin configuration can be mapped to
a {\em perfect matching}, which is a set of paths, made of
frustrated bonds, that join pairs of frustrated plaquettes. The
energy of the system is proportional to the total length (also
called weight) of these paths. Thus, finding a GS is equivalent to
finding a {\em minimum weighted perfect matching}.

To define the domain wall we use, as an example, the system shown
in Figure \ref{figure2} (generalization is straightforward). To
account for the free boundaries two plaquettes must be added, at
the top and the bottom (outside the lattice in Fig.~\ref{figure2})
\cite{Hartmann1}. The top (bottom) plaquette is called frustrated
if there is an odd number of frustrated bonds in the top (bottom)
boundary. The antiperiodic boundary condition changes only the
frustration of these two plaquettes, which in turn leads to a new
matching as GS. From the comparison of the two GSs a set of
contours can be defined, where each contour is formed by
alternating paths from each matching (see Fig.~\ref{figure2}). All
these contours will be closed (as the one joining plaquettes 8 to
11 in Fig.~\ref{figure2}), except for one that runs from the top
plaquette to the bottom one. This last contour is the domain wall
(it can be proved that the loops do not contribute to the energy
change).



\begin{figure} 
\includegraphics[width=6.5cm,clip=true]{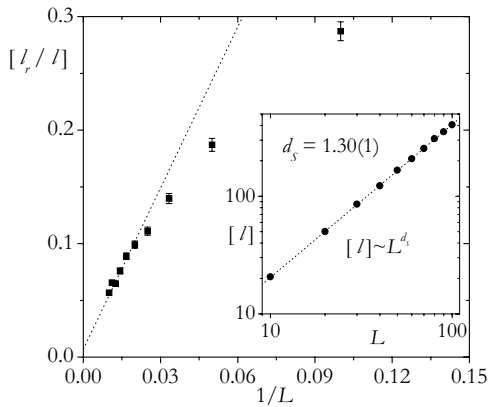}
\caption{\label{figure3} Scaling of ratio $\left[l_r / l \right]$
with $1/L$ for 2D. The inset shows the scaling of $\left[ l
\right]$ with $L$. Error bars are smaller than the symbols.
Averages have been taken over 1000 samples of each lattice size up
to $L=100$.}
\end{figure}

The results of simulations performed for several sample sizes are
shown in Fig.~\ref{figure3}, where both the average length of the
wall, $\left[l \right]$, and the average fraction of RL bonds in
the domain wall, $\left[l_r / l \right]$, are shown. As
anticipated, this fraction vanishes for large $L$ although a very
small, but nonvanishing value, cannot be ruled out. The average
length of the domain wall follows a scaling ($\left[l \right] \sim
L^{d_s}$) that allows us to find its fractal dimension, with
$d_s=1.30(1)$. Remarkably, this value coincides with the fractal
dimension reported \cite{Hartmann4} for the 2D EA model with a
Gaussian distribution of bonds.

In summary, we have studied the relevance of characterizing the
domain wall of the EA $\pm J$ spin glass by using topological
information of the GS. The defect energy on the RL in 3D shows a
behavior typical of a highly stable phase (similar to the 3D
ferromagnetic Ising model, but with $d_s>2$). On the other hand,
the FL shows a very different behavior, like a system in an
excited state. The total defect energy $\Delta E$ is the result of
these competitive and correlated effects (the whole system shows a
low stability with a small stiffness exponent).

The 2D case is very different. We have shown that the defect
energy avoids the RL, lying almost completely on FL (which
percolates in 2D). If we assume that the RL is the only structure
able to support a stable phase, then this behavior is compatible
with an unstable phase and a zero critical temperature. This
agrees with most recent studies (see e.g.~\cite{Hartmann5}).

In addition, for both dimensionalities we obtain the fractal
dimension of the domain wall by direct means. These values are in
good agreement with the ones reported in the literature for the EA
model with Gaussian distributed couplings.

We want to stress that our work is yet another indication that
separating the contributions of the backbone and its environment
provides relevant and non-trivial information about the nature of
the critical behavior of the $\pm J$ EA model. We believe that
this separation should also be important in the study of other
physical quantities. Moreover, this analysis can be applied to the
other systems (e.g. K-SAT problem \cite{Monasson}, Viana-Bray spin
glass \cite{Kanter}, etc.) that are known to have a backbone.

We are working to extend the concept of backbone to systems with
nondegenerated GSs, as happens with continuous distributions of
bonds. Results are still too preliminary to be reported here.


We thank S. Bustingorry and P. M. Gleiser for
helpful discussions and suggestions. FR and EEV thanks  Millennium
Scientific Iniciative (Chile) under contract P-02-054-F for
partial support. FR, FN and AJRP thanks Univ. Nac. de San Luis
(Argentina) under project 322000. We acknowledge support from
CONICET (Argentina) and FONDECYT (Chile) under projects PIP6294
and 1060317 and 7060300, respectively.


\begin{thebibliography}{99}
\bibitem{NGS} Because of the rapid growth of the number of the GSs, the average has been approximated using a large number of randomly sampled GSs, $N_{GS}$ (see Table I).
\end{thebibliography}
\end{document}